# Complex Economic Activities Concentrate in Large Cities

Pierre-Alexandre Balland[a,b], Cristian Jara-Figueroa[b], Sergio Petralia[c], Mathieu Steijn[a,d], David Rigby[e], César A. Hidalgo[b]

[a]Department of Human Geography and Spatial Planning, Utrecht University
[b]Collective Learning Group, The MIT Media Lab, Massachusetts Institute of Technology
[c]Department of Geography, London School of Economics
[d]Vrije Universiteit Amsterdam
[e]Department of Geography, University of California, Los Angeles

**Abstract**
Why do some economic activities agglomerate more than others? And, why does the agglomeration of some economic activities continue to increase despite recent developments in communication and transportation technologies? In this paper, we present evidence that complex economic activities concentrate more in large cities. We find this to be true for technologies, scientific publications, industries, and occupations. Using historical patent data, we show that the urban concentration of complex economic activities has been continuously increasing since 1850. These findings suggest that the increasing urban concentration of jobs and innovation might be a consequence of the growing complexity of the economy.

**Introduction**
In the year 2000, the San Francisco Bay Area produced more than 139 patents per 100,000 people, or 12% of all patenting activity in the United States. Fifteen years later, the Bay Area had more than doubled its per capita patenting output, generating more than 340 patents per 100,000 people. The Bay Area accounts for over 18% of all patenting activity in the U.S. This is more U.S. patents than any country in the world with the exception of Japan. But why is invention concentrated in places like the Bay Area? And why has this concentration increased so rapidly despite recent advances in communication and transportation technologies? Is the spatial concentration of economic activities a special characteristic of patents in high-tech industries, or is it a more general feature affecting all sectors of the economy?

In this paper we show that the more complex an economic activity is, the larger is its tendency to concentrate in large cities. We find this to be true for patents, research papers, industries, and occupations. Complex industries, such as biotech and semiconductors, exhibit a much greater degree of concentration in large cities than less complex industries such as apparel and furniture manufacturing. Using historical patent data, we show that the concentration in large cities of more complex inventions has increased continuously since at least 1850, while that of the least complex technologies has decreased since the 1970s.

The agglomeration of economic activities is considered a key ingredient of knowledge creation and economic growth (*1*, *2*). The standard hypotheses used to explain agglomeration is that firms care about sharing, learning, and matching (*3*). Firms seek locations where they can share inputs with other economic agents, learn from others, and match with the right employees. However, we still have much to learn about why some economic activities concentrate more than others.

Two recently developed strands of literature can shed some light on the factors that explain the variation of spatial concentration across activities: the literature on urban concentration (*4*) and the literature on economic complexity (*5–7*). The literature on urban concentration shows evidence that economic outputs increase faster than city size (they scale super-linearly) (*4*). This super-linear scaling is known to vary across economic activities (*8*, *9*), but it is not clear why some activities scale more super linearly than others. The literature on economic complexity has shown that economies engaged in more complex—more knowledge intense—activities are richer and grow faster (*5*, *10*). Here, we show that the spatial concentration of economic activities increases with their complexity.

But why would complex economic activities concentrate more in large cities? More complex economic processes require a deeper division of knowledge (*11*), and thus, operate more efficiently in large cities. This is consistent with the idea that knowledge complexity pushes individuals to narrow their expertise and specialize (*12*). This division of knowledge creates coordination costs that can be solved by the multiple interaction opportunities provided by cities (*13*, *14*).

Here, we validate this idea by showing that differences in the urban concentration of economic activities, as measured by their scaling exponents (*4*), are largely explained by differences in their level of complexity. That is, technologies that recombine more recent knowledge, research fields that involve larger scientific teams, industries that hire more educated workers, and occupations that require more years of education, are more concentrated in large cities than less complex technologies, research papers, industries and occupations. Moreover, using historical patent data going back to 1850, we show that the concentration of more complex forms of knowledge production has increased continuously over the last century and a half. These findings explain why some economic processes concentrate disproportionally in cities and contribute to our general understanding of the spatial organization of the economy.

**DATA**

We analyze the spatial distribution of patents, research papers, industries, and occupations in 353 Metropolitan Statistical Areas (MSAs) of the United States. For recent patents, we use the Patent Network Dataverse (*15*), providing longitude and latitude coordinates of inventor addresses for patents granted by the United States Patent and Trademark Office (USPTO) from 1975 to 2010. For historical patents (1850-1974), we use HistPat. HistPat was built using optically recognized and publicly-available documents from the USPTO, combining text-mining algorithms with statistical models to provide geographical information for historical patent documents (*16*). We disaggregate patents into 30 technologies as defined by the National Bureau of Economic Research (2-digit sub-categories)(*17*). For scientific papers, we use publication data from Elsevier's Scopus database covering the time period 1996–2008 (*8*, *18*). Publications are disaggregated into 23 scientific disciplines as defined by the Scopus classification (2-digit major thematic categories). For industries, we use 2015 GDP data from the Bureau of Economic Analysis to quantify the economic output of MSAs in 18 industries as defined by the North American Industry Classification System (2-digit NAICS). For occupations, we use 2015 employment statistics from the Bureau of Labor Statistics (BLS) disaggregated into 22 occupations according to the Standard Occupational Classification system (2-digit SOC). Population data originate in the US Census. See supplementary material (SM – section 1) for additional information.

**RESULTS**

Figure 1 shows the urban concentration of patents (Fig. 1 A), research papers (Fig 1 B), industries (Fig 1 C), and occupations (Fig 1 D) in the United States. Peaks are, respectively, proportional to the number of patents, the number of research papers, GDP, and the total employment of each metro area. In all four cases we find economic activities to be highly concentrated, especially in large cities. Figures 1 E-H characterize this urban concentration by showing the scaling laws followed by patents, research papers, industries, and occupations. Scaling-laws in cities are power-law relationships of the form $y \sim x^\beta$, where $x$ is the population of a city, $y$ is a measure of output (patents, papers, GDP, or jobs), and $\beta$ is the scaling exponent. In the case of patents (Figure 1 E), the number of patents granted to a city scales super-linearly with population with an exponent of $\beta$ =1.26. In the case of research papers, the number of papers published by authors in a metro area grows as the $\beta$ =1.54 power of that metro area's population. GDP on the other hand, grows as the $\beta$ =1.11 power of population, and total employment grows as the $\beta$ =1.04 power of the population in an MSA (these scaling exponents are in agreement with those reported in (*7*)).

Next, we repeat this exercise by studying the scaling laws followed by specific technologies, research areas, industries, and occupations. Figure 1 I shows the scaling laws followed by patents in "Computer, Hardware, and Software" and "Pipes & Joints." Patents in "Computer, Hardware, and Software" concentrate more in large cities ($\beta$ =1.57) than patents in "Pipes & Joints," which exhibit only a modest super-linear scaling ($\beta$ =1.1). Similarly, we observe large variations in the scaling coefficients of intuitively more and less complex research areas (Figure 1 J), industries (Figure 1 K), and occupations (Figure 1 L). Figures including each category of patents, papers, industries, and occupations are available in section 2 of the SM.

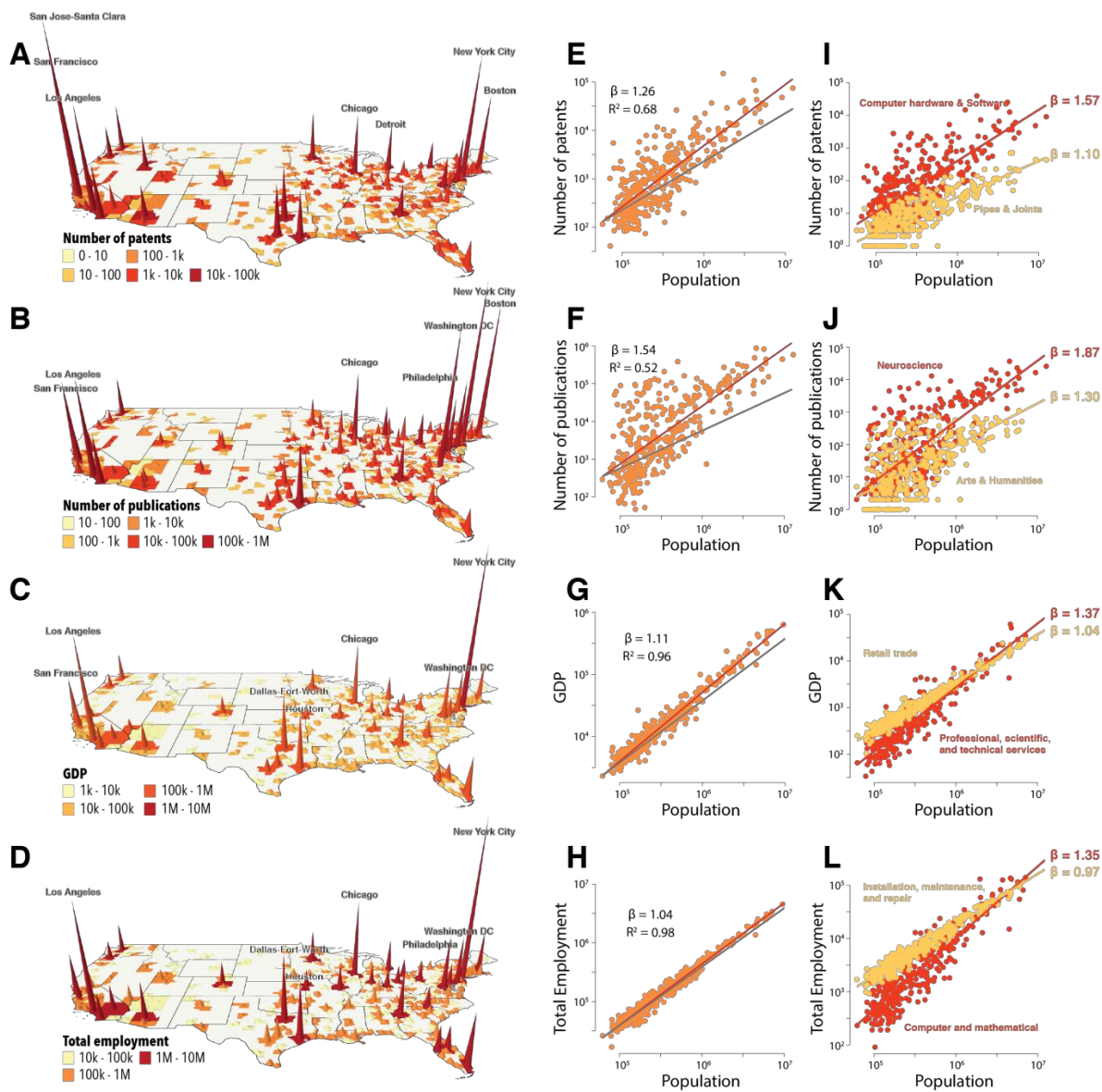

**Figure 1: Spatial concentration of economic activities.** Maps of the spatial concentration of **A** patents, **B** research papers, **C** GDP of industries, and **D** number of employees in U.S. MSAs. Scaling relations between the population of an MSA and **E** the total number of patents granted from 2000 to 2009, **F** the number of research papers published from 1996 to 2008, **G** the GDP of industries in 2015, and **H** the number of employees in 2015. Scaling relationships for pairs of economic activities with large differences in their scaling exponents: **I** patents in "computer, hardware, and software" and "pipes and joints", **J** Research papers in "Neuroscience" and "Arts and Humanities," **K** economic output (GDP) of "professional and scientific activities" and "retail trade", **L** employment in "computer and mathematical" occupations and in "installation, maintenance, and repair."

In Figure 2, we explain the urban concentration of economic activities using measures of their knowledge complexity. For technologies, we measure complexity using the vintage of the knowledge combined in the patent, measured as the average year of appearance of the sub-classes in which the patent makes a knowledge claim. This assumes patents that recombine more recent knowledge are—on average—more complex (*19*). For scientific fields, we use the average size of the team involved in a scientific publication. Team size is a direct interpretation of the idea that complex scientific activities require a finer division of knowledge (*20*). For industries, we use the average years of education of an industry's employees. For occupations, we use average years of education as a measure of the specialization required to participate in each activity. Because we compare the spatial concentration of economic activities with their economic complexity, we avoid using complexity measures that are derived from spatial information (*8*). For more information about these definitions and robustness analyses see section 3 of the SM.

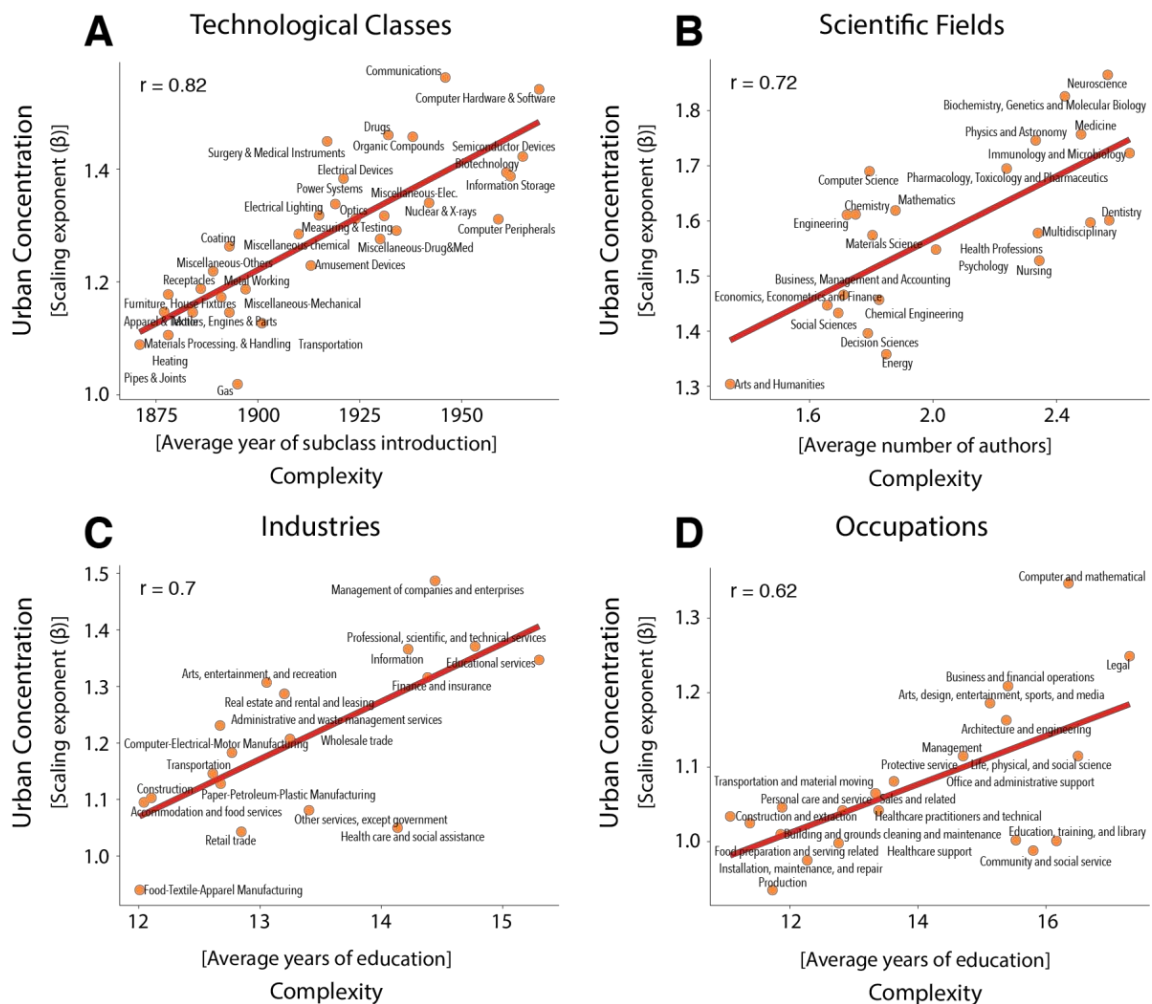

**Figure 2: Urban concentration increases with complexity.** Urban concentration of economic activities, as captured by the scaling exponent, increases with **A** The average year when the patent sub-classes were introduced; **B** the average number of authors in a publication in a field;

**C** the average years of education of the workers employed in an industry; and **D** the average years of education of workers within an occupational category.

Figures 2 A-D compare the urban concentration of economic activities with their respective scaling exponents. In all cases, we observe that the spatial concentration of economic processes increases with their knowledge complexity. For technologies, it increases with the recency of the combined sub-classes (Pearson's r = 0.82, $p<1\times10^{-3}$); for scientific fields, it increases with the average number of authors of a paper in a field (Pearson's r = 0.72, $p<1\times10^{-3}$); for industries, it increases with the years of education of the workers employed in that industry (Pearson's r = 0.70, $p<1\times10^{-3}$); and for occupations, it increases with the average years of education of the workers within that occupational category (Pearson's r = 0.62, $p<1\times10^{-3}$). In all four cases, the more complex the economic activity, the more super-linearly it scales with population, meaning that more complex economic activities concentrate more in large cities. We confirm the statistical significance of this relationship using regression analysis and a variety of alternative measures of spatial concentration and economic complexity (see section 3 of SM).

Next, we look at historical data to ask whether the concentration of economic activities has increased with the complexity of the economy? To explore this question, we use historical patent data, since it provides the longest time series (going back to 1850).

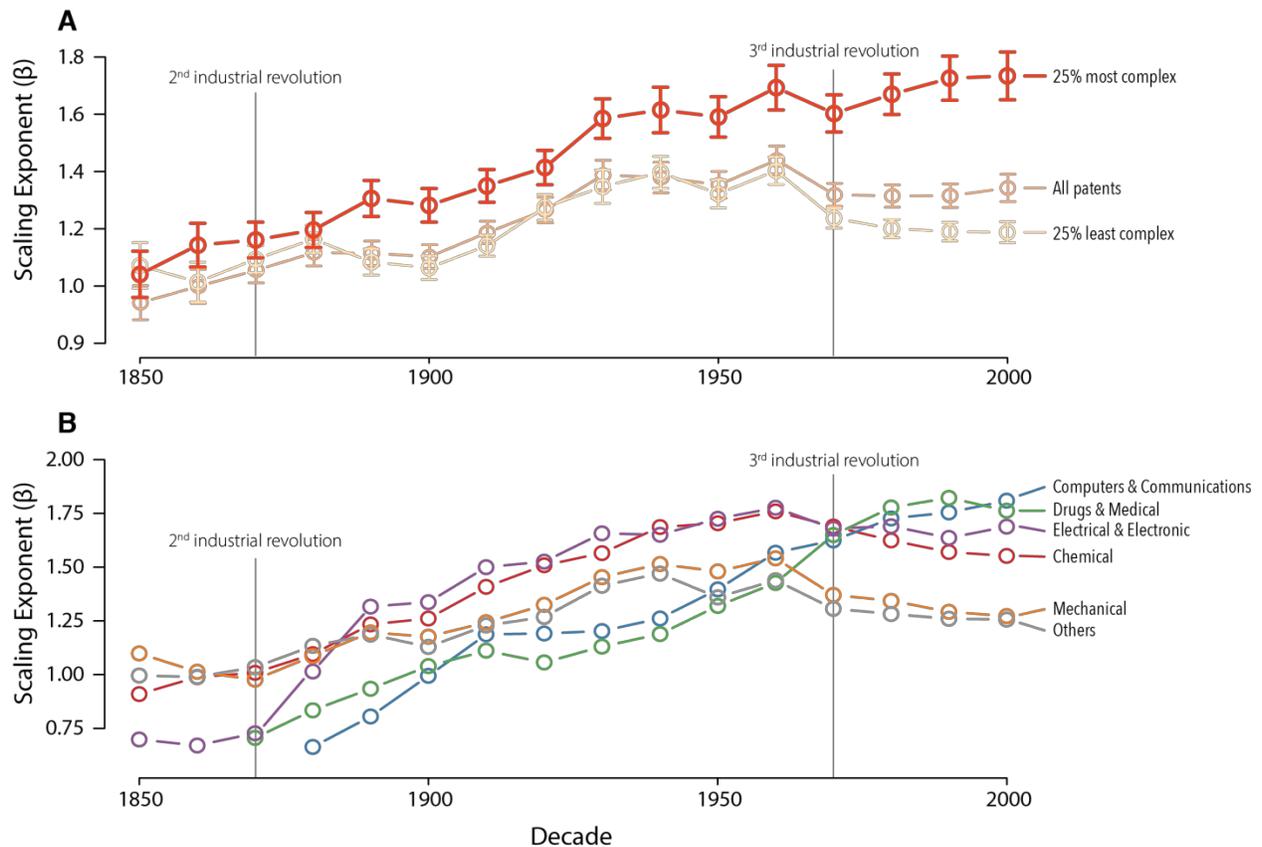

**Figure 3: Historical scaling of patenting activity. A** The scaling exponent of the top 25% most complex technologies increases throughout the observation period, while that of the bottom

25% of technologies based on complexity, peaks in 1960 and then decreases. The scaling exponent for all patents increases from 1850 to 1930, and then remains relatively stable until 2000. **B** The scaling exponent of the main six patent categories between 1850 and 2000.

Figure 3 A shows the scaling exponent observed for the top 25% most complex patents granted each decade between 1850 and 2010 (red line). It shows that the urban concentration of complex technologies, those that recombine newer knowledge, has continuously increased for the past 150 years and has accelerated with each industrial revolution. Starting with the second industrial revolution (1870), urban scaling of complex knowledge became increasingly super-linear, growing from a scaling exponent of $\beta \sim 1.15$ in 1870 to $\beta \sim 1.55$ by the 1930s. The urban concentration of the most complex patents then plateaued, and continued to increase after the 1970s I.T. revolution, reaching a scaling exponent of almost 1.8 in 2010. The least complex patents (light yellow line), on the other hand, have always been less concentrated than complex patents. After the 1970s, their urban concentration even started to decrease, with a scaling exponent falling to less than 1.2. The IT revolution has therefore been followed by an increasing concentration of the most complex technologies in cities, and a decreasing urban concentration of the least complex ones. Robustness analyses can be found in section 4 of the SM.

To further explore the evolution of the spatial concentration of patenting activity, we separate patents into their six main technological categories, as defined by the NBER: "Mechanical", "Chemical", "Electrical & Electronic", "Computers & Communication", "Drugs & Medical", and "Others". Figure 3 B shows the scaling exponent observed for each of these technological categories for each decade between 1850 and 2010. "Mechanical" and "Others" are the technologies that exhibit the highest scaling in the mid nineteenth century, meaning they are the ones most concentrated in large cities, with "Others" mostly composed of patents related to textiles during this period. The scaling exponent of these categories, however, does not grow substantially during the following decades, meaning that most of the rise in scaling observed after 1870 for all patents (Figure 3 A) can be attributed to an increase in the urban concentration of "Electrical & Electronic" patents. Starting in 1950, "Computers & Communications" and "Drugs & Medical" become increasingly more concentrated, reaching the highest scaling exponents observed for all categories. Together, these results show that the urban concentration of patenting activity exhibits a long-term cycle, rising during the heyday of the technologies developed, and then declining once technologies mature.

**DISCUSSION**

Why economic activities concentrate remains one of the longest standing puzzles in economic geography, urban science, and economic development. Yet, while there are many theories that can be used to explain the general tendency for economic activities to agglomerate (e.g. matching, learning, and sharing), we still need a better understanding of: (i) why some activities have a stronger tendency to agglomerate than others? And (ii), why economic activities continue to agglomerate despite recent advances in communication and transportation technologies?

Here we use differences in complexity to explain variations in the degree to which economic processes agglomerate. We argue that complex economic activities tend to be more concentrated in large urban areas because they require a deeper division of knowledge and labor. This also tells us that much of the (tacit) knowledge needed to perform these activities is embodied in social networks and that does not travel well through digital communication channels (*20*). The increase in agglomeration for more complex economic processes is measured in their respective scaling exponents. For patents, research papers, industries, and occupations, we find that the more complex, or more knowledge intensive the activity is, the more likely it is to exhibit super-linear scaling. Moreover, when we look at over a century of patenting activity in the U.S., we find that the dynamics of urban agglomeration are not static. On the contrary, the concentration of patenting activity in urban areas has increased during most of the last century and a half, especially during the second industrial revolution. During the IT revolution the concentration of the most and least complex activities diverged. The most complex technologies have reached unprecedented levels of urban agglomeration, while the least complex technologies experienced a decline in their agglomeration levels with the rise of communication technologies. This could explain why the world has become more flat for some activities (*22*), and more spiky for others (*23*).

The finding that more complex economic activities agglomerate more strongly has important implications for spatial inequality. If complexity and agglomeration cannot be divorced, the spatial inequality observed among large and small cities will continue to increase with future technological progress. This would happen as firms working in the complex economic activities that drive economic growth, such as pharma, artificial intelligence, and data services, continue to concentrate in a few large cities. Policymakers must recognize that the forces generating growth and innovation may be the same forces that are contributing to increasing spatial inequality.

**ACKNOWLEDGEMENTS**

We would like to thank Önder Nomaler, Koen Frenken, and Gaston Heimeriks for providing the data on scientific publications used in the main text, and Gregory Patience, Christian Patience, Bruno Blais, and Francois Bertrand for providing the data on the age of references listed on scientific publications. We also would like to thank Ron Boschma, Koen Frenken, Michael Storper, Allen J. Scott, Tom Broekel, and Frank Neffke for useful comments and suggestions. Financial support from the Regional Studies Association (RSA) through the Early Career Grant awarded to Pierre-Alexandre Balland is gratefully acknowledged. Cesar Hidalgo acknowledge support from the MIT Media Lab consortia, from MIT-Skoltech seed grant, and from the MIT-Masdar initiative.